\documentclass[aps,prb,twocolumn,10pt,superscriptaddress,showpacs]{revtex4-1}

\usepackage{bpchem}
\usepackage{graphicx}
\usepackage{color}
\begin{document}


\title{On entropy change measurements around first order phase transitions in caloric materials} 


\author{Luana Caron}
\email[]{caron@cpfs.mpg.de}
\affiliation{Max Planck Institute for Chemical Physics of Solids, N\"{o}thnitzer Str. 40, 01187 Dresden, Germany}
\author{Nguyen Ba Doan}
\affiliation{Univ. Grenoble Alpes, Inst NEEL, F-38000 Grenoble, France}
\affiliation{CNRS, Inst NEEL, F-38000 Grenoble, France}
\author{Laurent Ranno}
\affiliation{Univ. Grenoble Alpes, Inst NEEL, F-38000 Grenoble, France}
\affiliation{CNRS, Inst NEEL, F-38000 Grenoble, France}\date{\today}
\begin{abstract}
In this work we discuss the measurement protocols for indirect determination of the isothermal entropy change associated with first order phase transitions in caloric materials. The magneto-structural phase transitions giving rise to giant magnetocaloric effects in Cu-doped MnAs and FeRh  are used as case studies to exemplify how badly designed protocols may affect isothermal measurements and lead to incorrect entropy change estimations. Isothermal measurement protocols which allow correct assessment of the entropy change around first order phase transitions in both direct and inverse cases are presented.\end{abstract}
\maketitle 

Caloric effects are present in materials where the entropy may be modified by the application of an external field, resulting in a change in temperature \cite{Manosa:2013bh}. These may arise from the coupling between two different degrees of freedom: magnetocaloric (magneto-structural coupling) and electrocaloric (electro-structural coupling) effects; or from the strong response of a given variable, as is the case of the mechanocaloric effects where entropy and temperature changes are solely due to structural changes with little or no contribution from other degrees of freedom.
In the magnetocaloric and electrocaloric cases, magnetic or electric polarizations are strongly influenced by the application of magnetic and electric fields, respectively. Mechanocaloric effects result from the application of stresses and are called barocaloric in the isotropic case and elastocaloric in the uniaxial case. All caloric effects are characterized by the entropy and temperature changes caused by the application of an external field.

Entropy and temperature changes are particularly pronounced around first order phase transitions, making materials presenting large discontinuities in their polarization, volume or strain state as the result of the application of the corresponding conjugated field (the first order derivatives of the Gibbs free energy) the main focus of research in caloric effects.

For example, in magnetocaloric materials, a second order phase transition may give rise to entropy changes around 10 J/kgK in a 0-5 T magnetic field change, as is the case of the benchmark material, Gd\cite{Dankov:1998td}. Yet, a first order magneto-elastic phase transition yields twice as much in a 0-2 T magnetic field change, as is the case for  \BPChem{La(Fe\_{1-x}Si\_{x})\_{13}}\cite{fujita_itinerant-electron_2003} and \BPChem{Fe\_{2}P}-based\cite{dung_mixed_2011} materials. Therefore the latter are considerably more interesting than Gd for ferroic cooling applications, the chief driving force in the research for novel caloric materials.

However, discontinuous changes come with a high energetic cost which is manifest in thermal and field hysteresis (for isofield and isothermal measurements, respectively) as well as latent heat. This means that first order phase transitions are always partially irreversible and that not all of the entropy and temperature change observed in a material can be used in a cyclic manner. The issue of reversibility of both quantities upon cycling has been discussed by Basso et al.\cite{PhysRevB.85.014430} and more recently by Kaeswurm et al.\cite{Kaeswurm2016259} in the magnetocaloric case. 

However, hysteresis has a more immediate and fundamental consequence in the way we measure entropy change. 
In fact, entropy change, or entropy for that matter, cannot be measured. It can be obtained indirectly from specific heat measurements or from isothermal/isofield curves via the Maxwell equations:
\vspace{-0.2cm}
\[ \left( \frac{\partial S}{\partial Y_{i}} \right)_{T} = \left(\frac{\partial x_{i}}{\partial T}\right)_{Y_{i}} \]\vspace{-0.3cm}

where S is entropy, T temperature, $x_{i}$ and ${Y_{i}}$ are general displacements and fields, respectively.
 In the latter case, the thermodynamical history of the sample prior to measurement, i.e. the measurement protocol may have a profound influence on the calculated entropy change.

The controversy first arose within the magnetocalorics community, the oldest and better established of all caloric research lines. In the early 2000's, as the focus shifted from second to first order phase transitions, increasingly higher entropy changes derived from isothermal magnetization measurements started being reported, culminating in the extreme and unphysical report of the ``colossal" magnetocaloric effect in MnAs\cite{Gama:2004bz} and MnAs-based\cite{deCampos:2006fo} materials. MnAs is the textbook example of first order magneto-structural phase transition, showing extremely sharp magnetization changes and a pronounced thermal hysteresis. In these materials entropy changes above the magnetic limit $\Delta$S~=~Rln(2J+1) were reported, raising concerns about the validity of using the Maxwell equations. At the time, the Clausius-Clapeyron relation was put forward as an alternative\cite{PhysRevLett.83.2262}, and even a geometric argument\cite{liu_determination_2007} was proposed to remove the spurious peak. 

This controversy was resolved in a previous work by one of the present authors\cite{caron_determination_2009}. The Clausius-Clapeyron and Maxwell relations are  equivalent\cite{sun_comment_2000} and valid as long as they are applied to measurements performed between equilibrium states, which is exactly where the measurement protocols used at the time failed. A new protocol was proposed that took into account the thermodynamical history of a material and provided physical results. However, that work was left incomplete as the so-called loop process only addresses the transition from low to high magnetization in the conventional magnetocaloric case (where entropy decreases with increasing magnetic field). Moreover, this issue remains relevant to date as, in spite of the development of in-field differential scanning calorimeters (DSC), entropy change is still predominantly calculated from isothermal polarization measurements.

In this work we once more discuss how the isothermal measurement protocol may affect the calculated entropy change leading to spurious and non-physical results. The entropy change due to the magnetocaloric effect in Cu-doped MnAs and in FeRh are used as case studies for the direct and inverse caloric effects, respectively. Three isothermal measurement protocols are presented for each case: the commonly used second order protocol, and reset protocols for cooling and heating transitions for both direct and inverse MCE.

Preparation details of the \BPChem{Mn\_{0.99}Cu\_{0.01}As} sample used here are reported in a previous work and references therein\cite{caron_determination_2009}. A Ta (10~nm) / \BPChem{Fe\_{49.5}Rh\_{50.5}} (250~nm) / Ta (1~nm) trilayer was deposited by magnetron sputtering on a thermally oxidized Si substrate (\BPChem{SiO\_{2}} surface layer thickness = 100~nm). The polycrystalline layers were deposited at room temperature and subsequently ex-situ annealed at 923~K for 90~minutes. All magnetization measurements were performed using a Quantum Design MPMS XL, in fields up to 2~T in the case of \BPChem{Mn\_{0.99}Cu\_{0.01}As} and up to 5~T for \BPChem{Fe\_{49.5}Rh\_{50.5}} (applied parallel to the plane of the sample, no background was subtracted), in the reciprocating sample option (RSO).
\vspace{-0.4cm}
\section*{Direct MCE: \BPChem{M\MakeLowercase{n}\_{0.99}C\MakeLowercase{u}\_{0.01}A\MakeLowercase{s}} }
\vspace{-0.4cm}
As mentioned before, the measurement protocol initially used to measure isothermal magnetization for entropy change calculation was inherited from the second order case. In their 1999 paper\cite{pecharsky_magnetocaloric_1999-1} Pecharsky and Gschneidner give a detailed account of the indirect measurement methods for the magnetocaloric effect and explicitly show how the numeric integration of the Maxwell relation may be performed to obtain the entropy change. Albeit that no clear measurement protocol - other than isothermal - is presented, a consensus formed in the community where isotherms should be measured while increasing the magnetic field at temperature steps which were measured from low to high temperature. That does not represent a problem when there is only one value of magnetization for each temperature-field pair M(T,H) whatever the magneto-thermal history of your sample, as is the case of second order phase transitions. However, around first order phase transitions there will be two possible magnetization values for every M(T,H) pair, depending on the magneto-thermal history of the material. 
\begin{figure}[h]
\vspace{-0.2cm}
\centering
\includegraphics[width=0.47\textwidth]{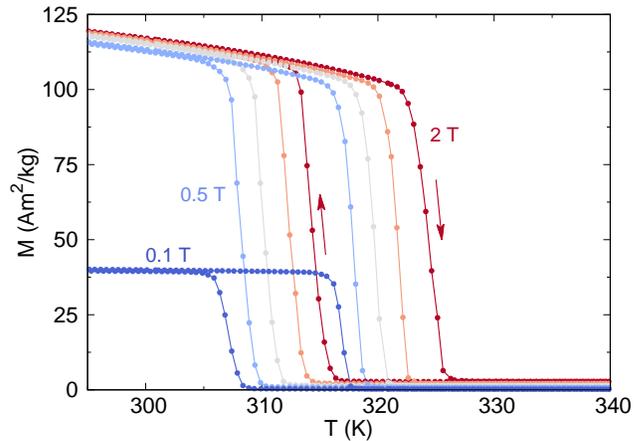}
\vspace{-0.4cm}
\caption{Temperature dependence of the magnetization at different applied fields. Between 0.5~T and 2~T curves are measured in 0.5~T steps.}
\label{MxTisoH}
\vspace{-0.4cm}
\end{figure}

\begin{figure}[h]
\vspace{-0.2cm}
\centering
\includegraphics[width=0.47\textwidth]{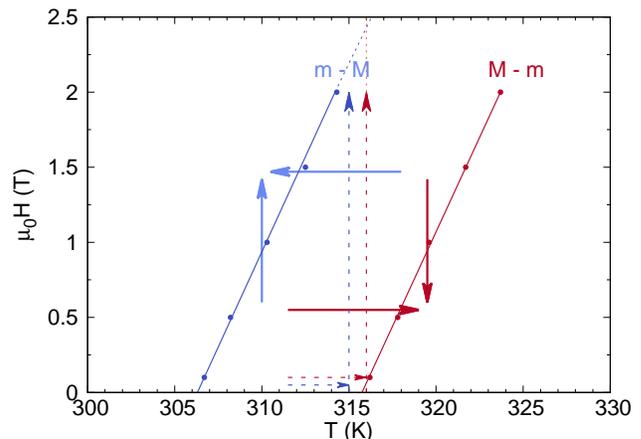}
\vspace{-0.4cm}
\caption{Critical field and temperature for both cooling/increasing field m - M and heating/decreasing field M - m transitions for \BPChem{Mn\_{0.99}Cu\_{0.01}As}.}
\label{TxH}
\vspace{-0.4cm}
\end{figure}

First let us analyze what goes wrong when applying a protocol devised for second order processes to a first order phase transition.
In a first order magnetic phase transition, be it direct or inverse, two transitions are observed: one from the high to the low magnetization state and another from the low to the high magnetization state. Let us call these transitions M - m and m - M, respectively, where M - m and m - M are obviously separated by the intrinsic thermal/field hysteresis. To make this analysis easier to grasp, we use \BPChem{Mn\_{0.99}Cu\_{0.01}As} as an example. This material shows a first order magneto-structural phase transition from a low temperature hexagonal ferromagnetic phase to a high temperature orthorhombic paramagnetic state, therefore displaying a direct MCE. In Fig.\ref{MxTisoH} are the isofield magnetization curves and in Fig. \ref{TxH} the critical field/temperature diagram derived from the isofield data in Fig.\ref{MxTisoH}. 

\begin{figure}[h]
\vspace{-0.2cm}
\centering
\includegraphics[width=0.47\textwidth]{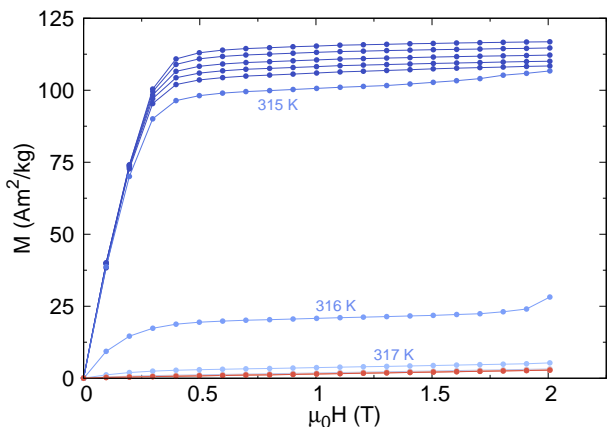}
\vspace{-0.6cm}
\caption{Magnetic isotherms measured using the protocol created to probe second order processes in the case of \BPChem{Mn\_{0.99}Cu\_{0.01}As}.}
\label{MxhisoT-P}
\vspace{-0.4cm}
\end{figure}

In the case of a direct transition such as that of \BPChem{Mn\_{0.99}Cu\_{0.01}As} the M - m transition (red line in Fig. \ref{TxH}) is crossed on increasing temperature or decreasing field and the m - M transition the reverse, decreasing temperature or increasing field. If the second order protocol is adopted to probe this first order process, isotherms are measured with increasing field from low to high temperature. Notice that, as long as T $<$ 316~K only the m - M transition may be crossed on increasing field. Since the material was already fully in the ferromagnetic (FM) state, the response will be that of simply rotating the magnetic domains and saturating the material, as indicated in Fig. \ref{TxH} by the blue dashed arrows and can be clearly seen in Fig. \ref{MxhisoT-P}. However, once the temperature is increased to 316~K at zero field, the sample will transform almost fully to the paramagnetic (PM) state as the M - m transition is crossed in temperature. Once the field is increased at 316~K, the FM fraction of the sample will be saturated giving rise to a plateau while the PM fraction can only transform to the FM phase when it crosses the m - M line at a field around 2.5~T as indicated by the red dashed arrows in Fig. \ref{TxH}. 

\begin{figure}[h]
\vspace{-0.2cm}
\centering
\includegraphics[width=0.47\textwidth]{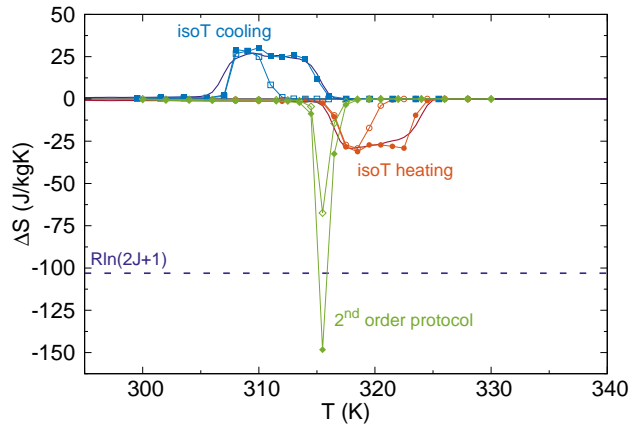}
\vspace{-0.5cm}
\caption{Entropy change for  0 - 1 T (open symbols) and 0 - 2 T (closed symbols) field changes using different protocols. Solid dark blue and red lines (without points) are calculated from the isofield curves in Fig. \ref{MxTisoH} for 0 - 2 T field change, on cooling and heating, respectively. The theoretical magnetic upper limit for the entropy given by Rln(2J+1) is indicated as a dashed line (J$_{eff}$ = 2 for the Mn atom).}
\label{dSxT2}
\vspace{-0.5cm}
\end{figure}

Notice that, increasing field implies that the m - M transition should be probed, but the second order protocol just followed measured two different processes: the field increase probes the m - M transition, while the temperature increase crosses the M - m transition. This has the effect of overestimating the isothermal entropy change calculated from this data as it concentrates the change in a very narrow temperature range. This gives rise, in extreme cases, to the so-called colossal MCE, which surpasses the theoretical magnetic upper limit for the entropy as can be observed in Fig. \ref{dSxT2} (green curve).

\begin{figure}[h]
\vspace{-0.2cm}
\centering
\includegraphics[width=0.47\textwidth]{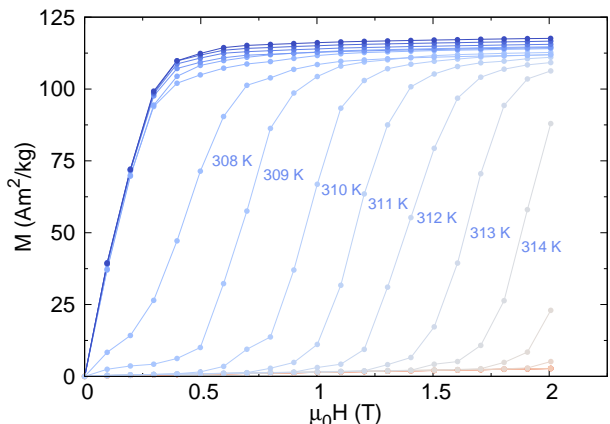}
\vspace{-0.6cm}
\caption{Magnetic isotherms measured using the loop protocol probing the m - M transition in the case of \BPChem{Mn\_{0.99}Cu\_{0.01}As}.}
\label{MxhisoT-loop0}
\vspace{-0.3cm}
\end{figure}

When measuring magnetic isotherms in order to calculate the magnetic entropy change, one must make sure to probe only one process: either M - m or m - M.
This can be achieved by bringing the material back to the starting point (or state) of the isothermal measurement once it has been performed. Thus one needs to devise a reversible thermodynamical cycle for each isotherm to be measured, literally going around the intrinsic irreversibility manifest in the thermal/field hysteresis.
In 2009 one of the authors of the present paper published the so-called loop process. The loop process resets the sample to the PM state in the case of a direct transition. This is done by increasing temperature at zero field far above the transition temperature in between isotherms and then measuring increasing field. This guarantees that only the m - M transition is crossed both in temperature and field. In the \BPChem{Mn\_{0.99}Cu\_{0.01}As} case, the sample was heated to 350~K at zero field. The results are in stark contrast to those obtained using the second order protocol (please compare Figs. \ref{MxhisoT-P} and \ref{MxhisoT-loop0}). A well developed field induced transition is observed which moves at a rate of 4~K/T as would be expected from the diagram presented in Fig. \ref{TxH} while the entropy change calculated from this measurement is well within the theoretical magnetic upper limit (see blue isoT cooling curve in Fig. \ref{dSxT2}).

\begin{figure}[h]
\vspace{-0.2cm}
\centering
\includegraphics[width=0.47\textwidth]{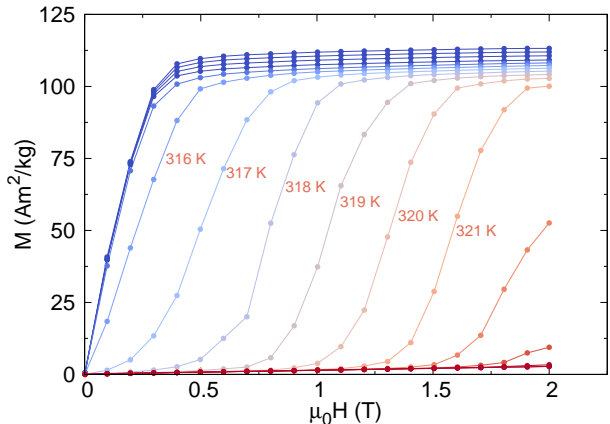}
\vspace{-0.6cm}
\caption{Magnetic isotherms measured using the loop protocol probing the M - m transition in the case of \BPChem{Mn\_{0.99}Cu\_{0.01}As}.}
\label{MxhisoT-loop2T}
\vspace{-0.4cm}
\end{figure}
A similar protocol must be used to probe the M - m transition. In this case the transition is crossed increasing temperature and/or decreasing field. Thus the sample must be taken well below the transition temperature, into the M state at the maximum field being used for measurements. In the case being presented here the \BPChem{Mn\_{0.99}Cu\_{0.01}As} sample was cooled to 280~K at 2~T in between isotherms, and only then heated to the next measurement temperature where magnetization was recorded upon decreasing magnetic field. The isotherms obtained in Fig. \ref{MxhisoT-loop0} are very similar to those in Fig. \ref{MxhisoT-loop2T}, shifted by 10~K due to thermal hysteresis, as is the entropy change (see orange isoT heating curve in Fig. \ref{dSxT2}).

Notice that, the second order measurement protocol, when applied to first order phase transitions, does not produce extra entropy. It was believed that in order to achieve the so-called colossal MCE, it was necessary to tap into other entropy reservoirs, such as the lattice entropy.\cite{ranke_analytical_2005} That is not the case and can be easily verified by calculating the area under the entropy change vs. temperature curve. For all three measurement protocols presented, the entropy content of the curves measured are, within error, the same. 

The entropy change was also calculated from isofield measurements (as presented in Fig. \ref{MxTisoH}). Isofield measurements have the advantage of always crossing the phase transition completely and univocally: it either crosses M - m or m - M. The entropy change calculated using isofield curves is shown as continuous lines (blue for cooling and orange for heating) on Fig. \ref{dSxT2} and is found to be in excellent agreement with those calculated from the isothermal curves measured using the reset protocols.\vspace{-0.4cm}
\section*{Inverse MCE: \BPChem{F\MakeLowercase{e}\_{49.5}R\MakeLowercase{h}\_{50.5}} }
\begin{figure}[h]
\vspace{-0.2cm}
\centering
\includegraphics[width=0.45\textwidth]{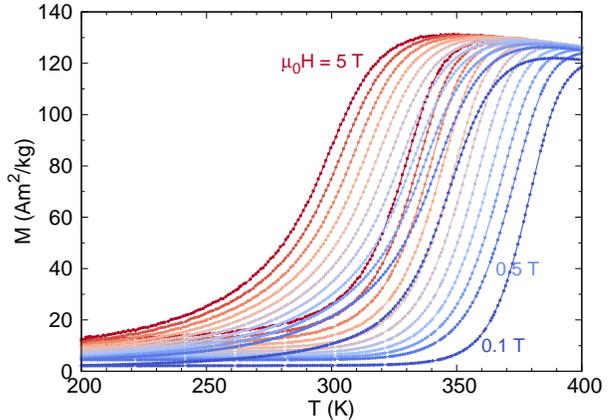}
\vspace{-0.4cm}
\caption{Temperature dependence of the magnetization in 0.1~T and from 0.5~T to 5~T in 0.5~T steps in \BPChem{Fe\_{49.5}Rh\_{50.5}} upon cooling and heating.}
\label{FeRh-MxTisoH}
\vspace{-0.4cm}
\end{figure}

\begin{figure}[h]
\vspace{-0.2cm}
\centering
\includegraphics[width=0.45\textwidth]{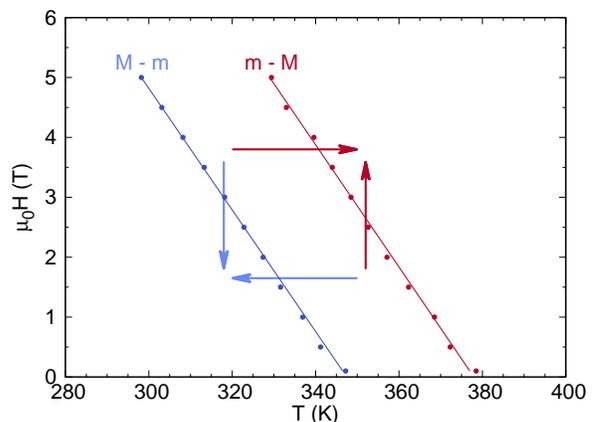}
\vspace{-0.4cm}
\caption{Critical field and temperature for both cooling/decreasing field M - m and heating/increasing field m - M transitions for \BPChem{Fe\_{49.5}Rh\_{50.5}} derived from the data in Fig. \ref{FeRh-MxTisoH}. Notice that the $dT_{C}/dH = -9.8 K/T$, which is above the value originally reported by Annaorazov et al..\cite{Annaorazov:1996ta}}
\label{TxH-inverse}
\vspace{-0.3cm}
\end{figure}

\vspace{-0.4cm}
The inverse transition stands as a very different case. Here field still has the role of taking the material from low to high magnetization state when increased and vice versa. But the temperature dependence of the magnetization has an inverse behavior compared to that of the direct transition: magnetization increases with increasing temperature.
Thus, to cross one or another phase transition the variables temperature and field must be changed in the same direction. To exemplify this case measurements on a thin film sample of \BPChem{Fe\_{49.5}Rh\_{50.5}} were performed. This material shows an isosymmetric phase transition between low temperature antiferromagnetic and high temperature ferromagnetic states\cite{Annaorazov:1996ta}. The temperature dependence of the magnetization under different applied magnetic fields are show in Fig. \ref{FeRh-MxTisoH} and the field dependence of the transition temperature in Fig.  \ref{TxH-inverse}. The transition from low to high magnetization i.e. m - M is crossed either increasing field and/or temperature (in red). Similarly to go from high to lower magnetization the magnetic field must be decreased and/or the temperature (in blue).

\begin{figure}[h]
\vspace{-0.2cm}
\centering
\includegraphics[width=0.45\textwidth]{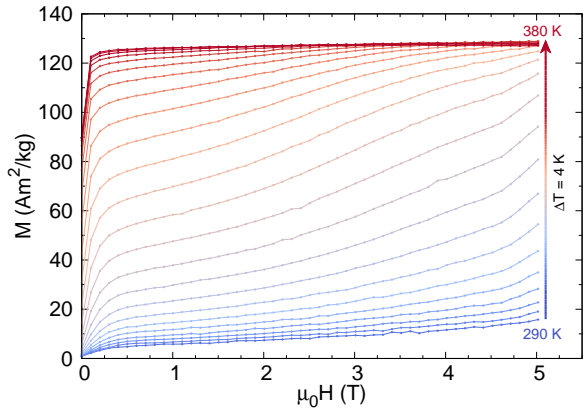}
\vspace{-0.4cm}
\caption{Magnetization isotherms measured using the 2$^{nd}$ order protocol for \BPChem{Fe\_{49.5}Rh\_{50.5}}.}
\label{FeRh-MxHisoT-2nd-5T}
\vspace{-0.5cm}
\end{figure}

\begin{figure}[h]
\centering
\vspace{-0.2cm}
\includegraphics[width=0.45\textwidth]{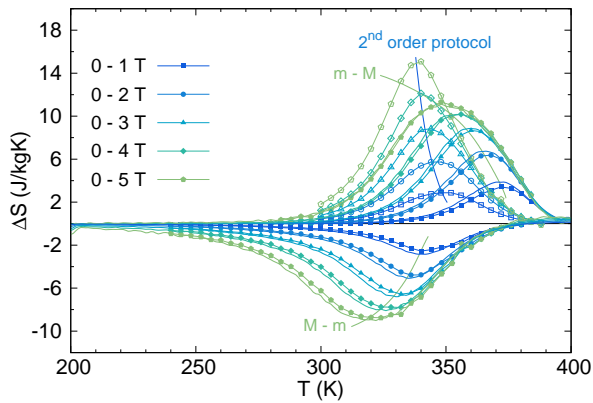}
\vspace{-0.4cm}
\caption{Temperature dependence of the entropy change at the first order phase transition of \BPChem{Fe\_{49.5}Rh\_{50.5}} computed from isotherms measured using the 2$^{nd}$ order protocol (open symbols), using reset protocols (closed symbols) for the m - M transition (positive entropy change) and the M - m transition (negative entropy change), and from the isofield curves presented in Fig. \ref{FeRh-MxTisoH} (continuous lines).}
\label{FeRh-dSxT}
\vspace{-0.3cm}
\end{figure}

As with the direct case, first the sample was measured using the standard second order protocol. Interestingly, the way temperature and magnetic field are changed in the second order protocol corresponds with crossing only the heating m - M transition in the inverse case. This may mislead one to think that the protocol correctly probes this transition without any problems with magneto-thermal history of the sample. However, in order to increase the magnetic field at a given temperature, the field must be decreased in between measurements, making the sample cross the M - m cooling transition every time as well and leading to the same problem observed around direct transitions.
In this case, since the transition itself is considerably broad, the effect is less clear than in the case of \BPChem{Mn\_{0.99}Cu\_{0.01}As}, and may easily induce error. The isotherms obtained using the 2$^{nd}$ order protocol are shown in Fig. \ref{FeRh-MxHisoT-2nd-5T} (isotherms for all protocols were measured at 4~K steps, as indicated in the figures). While a plateau is observed, changes are much more gradual and the entropy change calculated from this data falls within reasonable values for a giant MCE (see open symbols in Fig. \ref{FeRh-dSxT}). Only a more detailed analysis reveals that the entropy change maximum is not only overestimated, but also has its temperature dependence changed.


\begin{figure}[h]
\vspace{-0.2cm}
\centering
\includegraphics[width=0.45\textwidth]{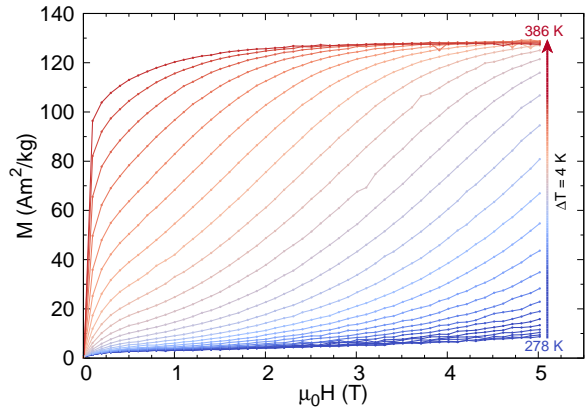}
\vspace{-0.4cm}
\caption{Magnetization isotherms measured using the reset protocol for the m - M (heating) transition for \BPChem{Fe\_{49.5}Rh\_{50.5}}. The sample was cycled down to 150~K at zero applied field between isotherms.}
\label{FeRh-MxHisoT-loop0}
\vspace{-0.5cm}
\end{figure}

Analogous to the direct case, in order to probe the m - M transition the isotherms must be measured with increasing field while the reset protocol must take the sample to the m state at low temperature at zero field. Thus the m - M line is always crossed either by increasing the field or by increasing temperature to the next measurement temperature after the reset. In this case the isotherms for the m - M transition were measured on increasing temperature and field while cycling the material well below the cooling transition, at 150~K and zero applied magnetic field in between measurements.
The isotherms obtained are shown in Fig. \ref{FeRh-MxHisoT-loop0}.

\begin{figure}[h]
\centering
\vspace{-0.2cm}
\includegraphics[width=0.45\textwidth]{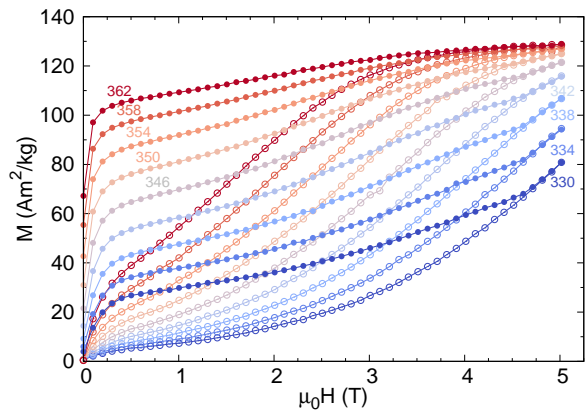}
\vspace{-0.4cm}
\caption{Magnetization isotherms measured using the 2$^{nd}$ order protocol (closed symbols) and the reset protocol for the m - M (heating) transition (open symbols) for \BPChem{Fe\_{49.5}Rh\_{50.5}}.}
\label{FeRh-MxHisoT-comp}
\vspace{-0.3cm}
\end{figure}

 Here the plateau found in Fig. \ref{FeRh-MxHisoT-2nd-5T} for the 2$^{nd}$ order protocol is no longer observed and the entropy change (positive entropy change curves represented using closed symbols in Fig. \ref{FeRh-dSxT}) shows a completely different temperature dependence as well as a lower maximum. The differences in these two processes can only be appreciated when curves at the same temperature for both protocols are shown together. In Fig. \ref{FeRh-MxHisoT-comp} nine isotherms are presented for both protocols, closed symbols are used for the 2$^{nd}$ order protocol while open symbols show isotherms measured using the reset protocol for the m - M (or heating) transition.

\begin{figure}[h]
\vspace{-0.2cm}
\centering
\includegraphics[width=0.45\textwidth]{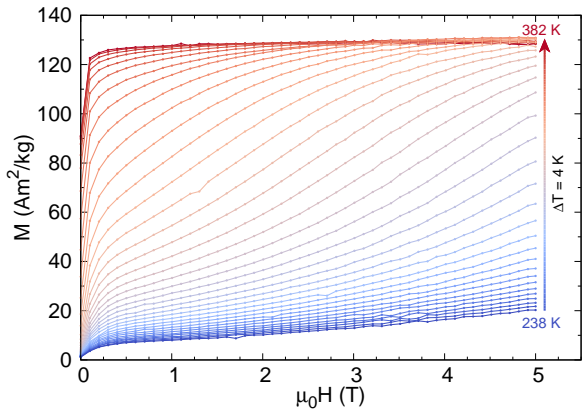}
\vspace{-0.4cm}
\caption{Magnetization isotherms measured using the reset protocol for the M - m (cooling) transition for \BPChem{Fe\_{49.5}Rh\_{50.5}}. The sample was cycled up to 390~K at 5~T between isotherms.}
\label{FeRh-MxHisoT-loop5T}
\vspace{-0.3cm}
\end{figure}

A similar logic is used to design the reset protocol for the the M - m transition, which must be measured with decreasing field steps. The reset protocol must take the sample above the transition at the maximum field being used for measurements to the point of highest magnetization and only then cooled, at the same maximum field, to the next measurement temperature. This guarantees the M - m transition alone is crossed: when decreasing the field to record the isotherm and/or when decreasing the temperature after the temperature loop at maximum field above the transition. For \BPChem{Fe\_{49.5}Rh\_{50.5}} the sample was cycled at 5~T up to  390~K between isotherms. These isotherms are shown in Fig. \ref{FeRh-MxHisoT-loop5T}, where well-developed metamagnetic phase transitions can be observed as in the case of the measurement of the m - M transition using the reset protocol shown in Fig. \ref{FeRh-MxHisoT-loop0}. The entropy change is shown in Fig. \ref{FeRh-dSxT} (closed symbols, negative entropy change curves). The maximum entropy change here is lower than for the heating transition as the transition is slightly broader, and is shifted in temperature by thermal hysteresis. 

Finally, the entropy change curves computed from isotherms measured using the reset protocols are in excellent agreement with the entropy change computed from the isofield measurements shown in Fig. \ref{FeRh-MxTisoH} (represented as continuous lines on Fig. \ref{FeRh-dSxT}), as expected.

Albeit simple, these protocols ensure that phase transitions are crossed univocally during isothermal measurements, in both direct and inverse cases. In this context, it is worthwhile to mention that isofield and properly reset isothermal magnetization measurements both probe equilibrium states and can be used as input to the Maxwell relation for the calculation of the entropy change. In the case of magnetocaloric materials isofield measurements have the advantage of always probing the phase transition correctly, and are found to be in excellent agreement with isothermal measurements (see Figs. \ref{dSxT2} and \ref{FeRh-dSxT}). However, not all caloric materials can be measured at a constant applied field. That is the case of electrocaloric materials where high electric fields would have to be sustained long enough for the temperature to be swept up and down, which may result in voltage breakdown.

Furthermore, by determining how isothermal measurements should be performed, these protocols are a first step towards standardizing entropy change calculation for caloric effects. Isothermal measurements performed using the correct protocol should produce reliable results that can be easily and readily compared in literature, a crucial issue for the development of these materials towards applications. 

In summary we have discussed the issue of isothermal measurements for the calculation of entropy changes using the Maxwell relation around first order phase transitions. Magnetocaloric Cu-doped MnAs and \BPChem{Fe\_{49.5}Rh\_{50.5}} were used as case studies of direct and inverse first order phase transitions, respectively, to exemplify the reset protocols necessary for both heating and cooling transitions.For the first time measurement protocols for heating and cooling in both direct and inverse transition cases have been created and demonstrated. Moreover these protocols apply to all caloric effects, being specially relevant for electrocaloric materials. This work not only aims at discussing and showing how to probe first order phase transitions correctly but also at standardizing the entropy change determination from isothermal measurements, a crucial step towards application of caloric materials. 
\vspace{-0.4cm}
\begin{acknowledgments}
\vspace{-0.4cm}
The FeRh film used in this study was developed in the framework of the European community's FP7 project No. 310748 (DRREAM).
\end{acknowledgments}%

\end{document}